\documentclass[aps,pra,10pt,twocolumn,superscriptaddress,longbibliography,showpacs,floatfix]{revtex4-1}
\usepackage{color}
\usepackage{amstext,amsmath,amssymb}
\usepackage{graphicx}
\usepackage[colorlinks,bookmarks=false,citecolor=darkblue,linkcolor=red,urlcolor=blue]{hyperref}

\definecolor{darkred}{rgb}{0.7,0.0,0.0}

\definecolor{darkblue}{rgb}{0,0.02,0.45}

\definecolor{darkgreen}{rgb}{0.02,0.45,0.0}

\definecolor{violet}{rgb}{0.8,0.2,0.6}

\newcommand{\eff}{\rm eff}

\begin{document}

\title{Persistent low-temperature spin dynamics in mixed-valence iridate Ba$_{3}$InIr$_{2}$O$_{9}$}

\author{Tusharkanti Dey}
\email[Email: ]{tusdey@gmail.com}
\affiliation{Experimental Physics VI, Center for Electronic Correlations and Magnetism,
University of Augsburg, 86159 Augsburg, Germany}

\author{M. Majumder}
\affiliation{Experimental Physics VI, Center for Electronic Correlations and Magnetism,
University of Augsburg, 86159 Augsburg, Germany}

\author{J. C. Orain}
\affiliation{Laboratory for Muon Spin Spectroscopy, Paul Scherrer Institut, 5232
Villigen PSI, Switzerland}

\author{A. Senyshyn}
\affiliation{Forschungsneutronenquelle Heinz Maier-Leibnitz (FRM II), Technische
Universit\"{a}t M\"{u}nchen, 85747 Garching, Germany}

\author{M. Prinz-Zwick}
\affiliation{Experimental Physics V, Center for Electronic Correlations and Magnetism,
University of Augsburg, 86159 Augsburg, Germany}

\author{S. Bachus}
\affiliation{Experimental Physics VI, Center for Electronic Correlations and Magnetism,
University of Augsburg, 86159 Augsburg, Germany}

\author{Y. Tokiwa}
\affiliation{Experimental Physics VI, Center for Electronic Correlations and Magnetism,
University of Augsburg, 86159 Augsburg, Germany}

\author{F. Bert}
\affiliation{Laboratoire de Physique des Solides, CNRS, Univ. Paris-Sud, Universit$\acute{e}$
Paris-Saclay, 91405 Orsay Cedex, France}

\author{P. Khuntia}
\affiliation{Laboratoire de Physique des Solides, CNRS, Univ. Paris-Sud, Universit$\acute{e}$
Paris-Saclay, 91405 Orsay Cedex, France}

\author{N. B\"{u}ttgen}
\affiliation{Experimental Physics V, Center for Electronic Correlations and Magnetism,
University of Augsburg, 86159 Augsburg, Germany}

\author{A. A. Tsirlin}
\email[Email: ]{altsirlin@gmail.com}
\affiliation{Experimental Physics VI, Center for Electronic Correlations and Magnetism,
University of Augsburg, 86159 Augsburg, Germany}

\author{P. Gegenwart}
\email[Email: ]{philipp.gegenwart@physik.uni-augsburg.de}
\affiliation{Experimental Physics VI, Center for Electronic Correlations and Magnetism,
University of Augsburg, 86159 Augsburg, Germany}

%\date{\today}
\begin{abstract}
Using thermodynamic measurements, neutron diffraction, nuclear magnetic resonance, and muon spin relaxation, we establish putative quantum spin liquid behavior in Ba$_3$InIr$_2$O$_9$, where unpaired electrons are localized on mixed-valence Ir$_2$O$_9$ dimers with Ir$^{4.5+}$ ions. Despite the antiferromagnetic Curie-Weiss temperature on the order of 10\,K, neither long-range magnetic order nor spin freezing are observed down to at least 20\,mK, such that spins are short-range-correlated and dynamic over nearly three decades in temperature. Quadratic power-law behavior of both spin-lattice relaxation rate and specific heat indicates gapless nature of the ground state. We envisage that this exotic behavior may be related to an unprecedented combination of the triangular and buckled honeycomb geometries of nearest-neighbor exchange couplings in the mixed-valence setting.
\end{abstract}

%\pacs{75.10.Kt, 76.75.+i, 76.60.-k, 75.10.Jm}

\maketitle
\section{Introduction}
Frustrated magnets host multiple exotic states, including quantum spin liquids (QSLs). In a QSL, spins are strongly correlated, but quantum fluctuations prevent them from long-range ordering~\cite{balents2010,savary2017}. The initial (and subsequently rebutted) proposal of the QSL resonating-valence-bond state on the triangular lattice of Heisenberg spins~\cite{anderson1973} was followed by similar proposals for several other isotropic (Heisenberg) \mbox{spin-$\frac12$} models, where the formation of QSLs is now established~\cite{yan2011,mishmash2013,kaneko2014,hu2015,iqbal2016}. A few candidate QSL materials proposed over the last decade bear key experimental signatures of this exotic state, including persistent spin dynamics and the absence of long-range order within the experimentally accessible temperature range~\cite{norman2016,fak2012,clark2013,balz2016}.

More recently, QSL states in anisotropic magnets have been explored. Here, Kitaev model with anisotropic interactions on the honeycomb lattice~\cite{kitaev2006} offers an exact solution for the QSL. Real-world manifestations of the Kitaev physics are found in compounds of $4d$ and $5d$ transition metals~\cite{jackeli2009}, where large spin-orbit coupling triggers strong intersite magnetic anisotropy. However, none of the Kitaev materials reported to date host the QSL ground state in zero field, and long-range order typically sets in at low temperatures, owing to substantial interactions beyond the Kitaev terms~\cite{winter2016,winter2017}. 

%%%%%%%%%%%%%%%%%%%%%%%%%%%%%%%%%%%%%%%%%%%%%%%%%%%%%%%%%%%%%%%%%
\begin{figure*}
\includegraphics[scale=0.6]{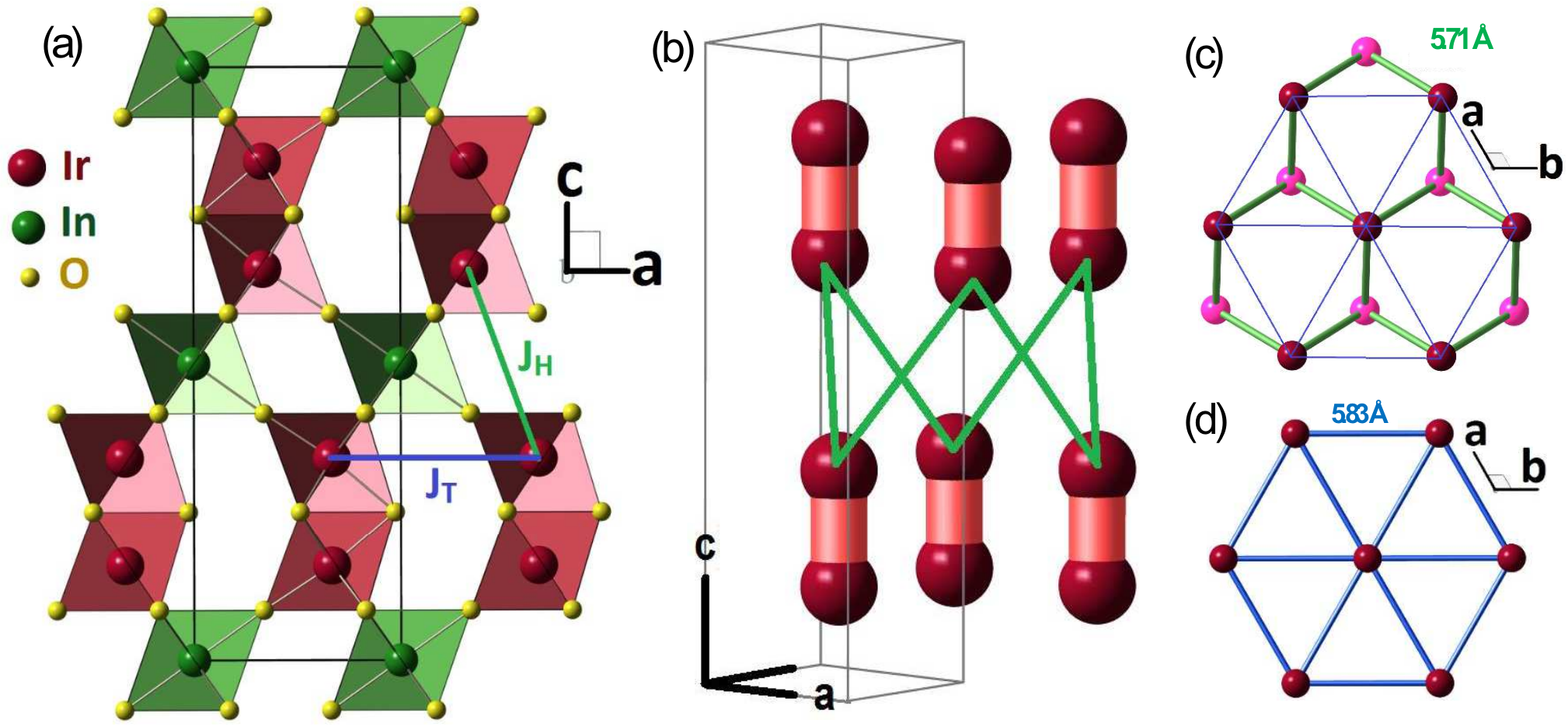}
\caption{\label{fig:Structure}
(a) The crystal structure of Ba$_3$InIr$_2$O$_9$ comprising the IrO$_{6}$ and InO$_{6}$ octahedra shown in red and green colors, respectively. The Ba atoms are omitted for clarity. (b) Magnetic moments are localized within the Ir dimers shown as dumbbells. The interplane couplings $J_H$ are shown as green lines. (c) The interplane coupling $J_H$ forming the buckled honeycomb spin lattice. (d) The intra-plane coupling $J_T$ forming the triangular spin lattice. 
}
\end{figure*}
%%%%%%%%%%%%%%%%%%%%%%%%%%%%%%%%%%%%%%%%%%%%%%%%%%%%%%%%%%%%%%%%%
In this paper, we propose an alternative strategy and search for QSL states in the family of mixed-valence $5d$ oxides, where unpaired electrons are localized on dimers of the Ir atoms. This should facilitate access to hitherto unexplored local electronic states~\cite{streltsov2016} and new regimes of anisotropic exchange interactions. Specifically, we report low-temperature magnetic behavior of the mixed-valence iridate Ba$_3$InIr$_2$O$_9$ as a QSL candidate, and confirm its persistent spin dynamics as well as the absence of long-range magnetic order down to at least 20\,mK. We further identify quadratic-like power-law behavior of both specific heat and spin-lattice relaxation rate, and compare these observations to existing theoretical results on QSLs. 

Ba$_3$InIr$_2$O$_9$ belongs to the family of hexagonal perovskites A$_3$BM$_2$O$_9$. Their structures comprise single BO$_6$ octahedra and M$_2$O$_9$ dimers of two face-sharing MO$_6$ octahedra (Fig.~\ref{fig:Structure}a). When magnetic ion occupies the B site, triangular interaction geometry is formed, as in Ba$_3$CoSb$_2$O$_9$, which is arguably the best model spin-$\frac12$ antiferromagnet on the triangular lattice~\cite{shirata2012,zhou2012,ma2016}. Placing a $4d$ or $5d$ ion into the B site or into one of the M sites could give rise to a triangular system with leading Kitaev interactions~\cite{becker2015,catuneanu2015}, but experimental implementation of this idea is hindered by the strong B/M site mixing that occurs, e.g., in Ba$_3$IrTi$_2$O$_9$~\cite{dey2012,kumar2016,lee2017}. Alternatively, Ir could be introduced into both M sites, while keeping the B site non-magnetic, but for integer valence of M such a spin dimer would simply condense into a non-magnetic singlet~\cite{darriet1976,darriet1983}. Mixed-valence systems with both M sites occupied by a magnetic $5d$ ion are possible too~\cite{doi2004,sakamoto2006,dey2013,dey2014}, and seem to be more promising for finding a QSL, because unpaired electrons localized on the dimers appear. Note that such \textit{mixed-valence dimers} with an unpaired electron delocalized between the two Ir$^{4.5+}$ ions are very different from more conventional \textit{spin dimers} formed by two magnetic ions holding one unpaired electron each.

Guided by this idea, we synthesized polycrystalline samples of Ba$_3$InIr$_2$O$_9$. From our detailed study using  neutron diffraction, magnetization and specific heat measurements, muon spin relaxation ($\mu$SR), and nuclear magnetic resonance (NMR) we establish a gapless and, potentially, spin-liquid ground state in Ba$_3$InIr$_2$O$_9$. 

\section{Experimental Details}

Polycrystalline samples of Ba$_{3}$InIr$_{2}$O$_{9}$ were prepared by a conventional solid-state reaction method ~\cite{sakamoto2006}. Stoichiometric amounts of high-purity BaCO$_{3}$, In$_{2}$O$_{3}$, and Ir metal powder were mixed thoroughly, pressed into pellets, and calcined at $900$\,$^{\circ}$C for $12$\,h. Further, the pellet was crushed into powder, mixed well, pelletized, and fired at $1300$\,$^{\circ}$C for $4$\,days with several intermediate grindings. Neutron diffraction data were collected at the high-resolution instrument SPODI ~\cite{spodi} at FRM-II (TU Munich) using the wavelength of 1.55\,\r A. Jana2006 software ~\cite{jana2006} was used for structure refinement. 

Magnetization measurements were carried out in a Quantum Design $5$\,T SQUID magnetometer and in a Quantum Design PPMS $14$\,T equipped with the vibrating sample magnetometer in the temperature range $2-350$\,K. Additional high-temperature data extending up to 650\,K were collected in the SQUID magnetometer using powder sample enclosed in a thin-walled quartz tube. 

Heat-capacity measurements in the temperature range $0.4-200$\,K were performed in a Quantum Design PPMS using the $^3$He insert. Low-temperature measurements in the $0.08-1$\,K range were performed using a quasi-adiabatic heat pulse method, adapted to a dilution refrigerator. 

%%%%%%%%%%%%%%%%%%%%%%%%%%%%%%%%%%%%%%%%%%%%%%%%%%%%%%%%%%%%%%%%%
\begin{figure}
\includegraphics[width=0.48\textwidth]{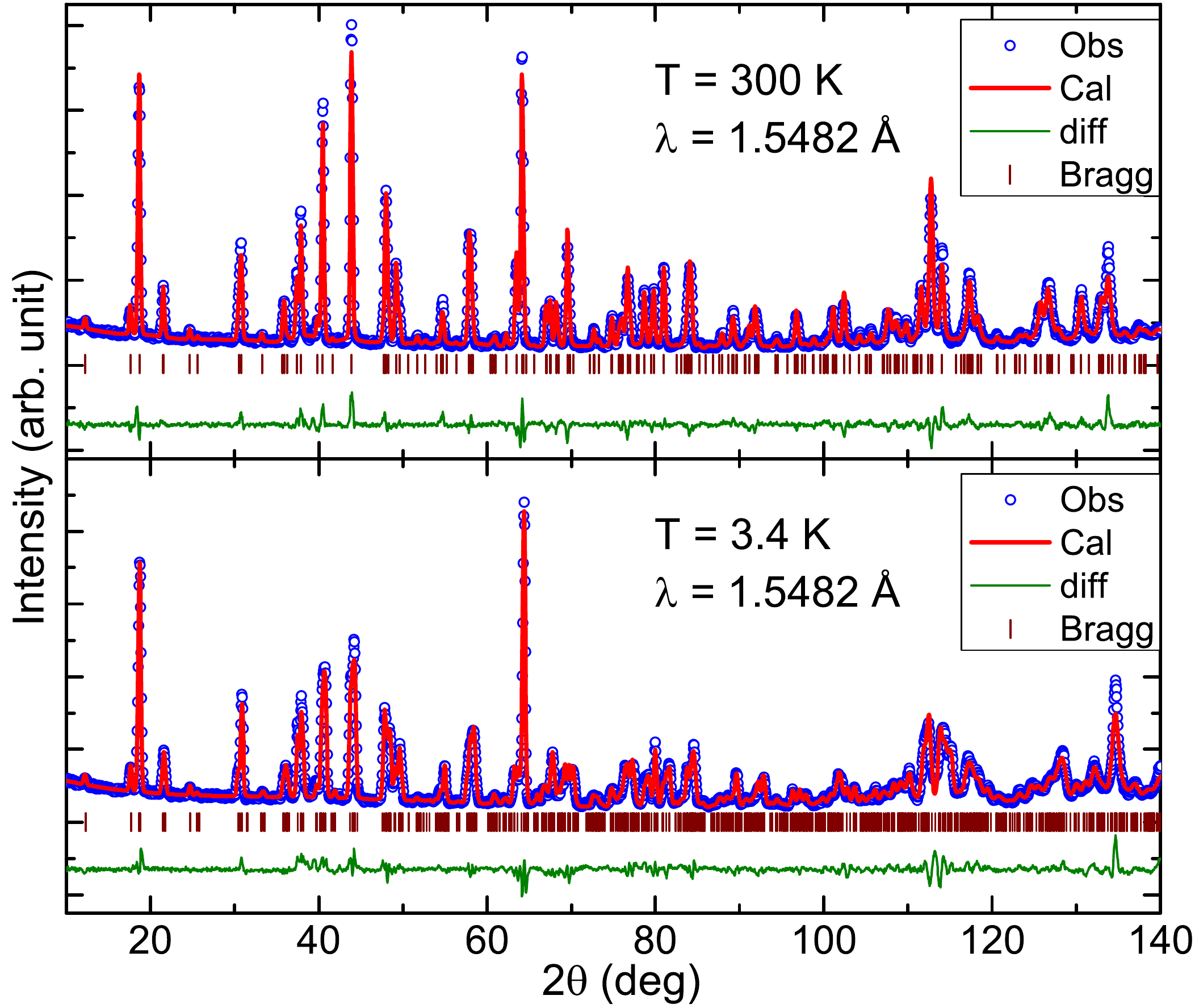}
\caption{\label{fig:NeutronDiff}
Rietveld refinement of neutron diffraction data at room temperature (top) and at 3.4\,K (bottom).
}
\end{figure}
%%%%%%%%%%%%%%%%%%%%%%%%%%%%%%%%%%%%%%%%%%%%%%%%%%%%%%%%%%%%%%%%%

$\mu$SR experiments were done on two different spectrometers at the Paul Scherrer Institute (Switzerland), LTF for temperatures from 20\,mK up to 750\,mK and Dolly for temperatures from 250\,mK up to 200\,K. For the Dolly experiment, about 300\,mg of the polycrystalline sample was mounted on a thin copper plate inside the $^{3}$He cryostat. In order to ensure good thermal contact, we glued the sample with GE varnish. We used the Veto mode, which allowed to get rid of the background signal from the sample holder. Therefore, the acquired signal is due to muons that stopped inside the sample. For the LTF experiment we used the same sample, again glued with GE on a silver plate. By comparing the results obtained between 250\,mK and 750\,mK on both the spectrometers, we were able to get rid of the experimental background on LTF.

$^{115}$In nuclear magnetic resonance (NMR) experiments were carried out with our home-built spectrometer with the dilution-fridge insert. The measurements are performed in the field-sweep mode at a fixed frequency of $70$\,MHz down to $24$\,mK. The spin echo intensity was obtained by integrating over the spin echo in the time domain. The final spectrum is constructed by plotting the spin-echo intensity as a function of the applied field. 

\section{Results}
\subsection{Crystal structure}

Rietveld refinement of room-temperature neutron diffraction data (Fig.~\ref{fig:NeutronDiff}) confirms hexagonal crystal structure ($P6_3/mmc$). However, at 3.4\,K peak splitting of the 203 and 204 reflections, as well as a visible broadening of other peaks, indicated that the symmetry is reduced to monoclinic. The 3.4\,K data were refined in the $C2/c$ space group similar to other hexagonal perovskites~\cite{doi2004,sakamoto2006,ling2010}. Both hexagonal and monoclinic structures feature a single crystallographic position of Ir (Tables~\ref{tab:structure} and~\ref{tab:structure2}) suggesting the true intermediate-valence Ir$^{4.5+}$ state. This is different from, e.g., Ba$_5$AlIr$_2$O$_{11}$, where two sites of the dimer belong to two different crystallographic positions, thus making possible charge re-distribution within the dimer~\cite{terzic2015}. 

In contrast to Ba$_3$IrTi$_2$O$_9$ with its $35-40$\% site mixing~\cite{dey2012,lee2017}, our mixed-valence Ba$_3$InIr$_2$O$_9$ shows high degree of structural order. We were able to obtain reasonable atomic displacement parameters in the fully ordered models of both hexagonal and monoclinic structures (Tables~\ref{tab:structure} and~\ref{tab:structure2}). On the other hand, deliberate admixing of In into the Ir position and vice versa leads to a marginal reduction in the refinement residuals and 2.8(5)\% site mixing~\footnote{We introduced the site mixing and constrained the overall composition to Ba$_3$InIr$_2$O$_9$. This leads to $R_I=0.038$ and $R_p=0.050$ compared to $R_I=0.039$ and $R_p=0.051$ for the fully ordered model (room-temperature data).}. We believe that the diffraction data alone may not distinguish between the fully ordered structure and the weak site mixing scenario. Further studies, such as direct imaging with high-resolution electron microscopy, would be useful to resolve this issue.

%%%%%%%%%%%%%%%%%%%%%%%%%%%%%%%%%%%%%%%%%%%%%%%%%%%%%%%%%%%%%%%%%
\begin{table}
\caption{\label{tab:structure}
Refined atomic positions for Ba$_3$InIr$_2$O$_9$ at 300\,K. The $U_{\rm iso}$ are isotropic atomic displacement parameters (in 10$^{-2}$\,\r A$^2$). Lattice parameters: $a=5.8316(1)$\,\r A, $c=14.4877(4)$\,\r A, $P6_3/mmc$, $R_I=0.039$, $R_p=0.051$.
}
\begin{ruledtabular}
\begin{tabular}{cccccc}
 & & $x/a$ & $y/b$ & $z/c$ & $U_{\rm iso}$ \\
Ba1 & $2b$ & 0 & 0 & 0.25 & 1.3(1) \\
Ba2 & $4f$ & $\frac13$ & $\frac23$ & 0.9103(2) & 0.9(1) \\
In  & $2a$ & 0 & 0 & 0 & 0.4(1) \\
Ir  & $4f$ & $\frac13$ & $\frac23$ & 0.1590(1) & 1.29(3) \\
O1  & $6h$ & 0.4867(2) & 0.5133(2) & 0.25 & 1.40(5) \\
O2 & $12k$ & 0.1715(2) & 0.3430(4) & 0.0841(1) & 2.1(1) \\
\end{tabular}
\end{ruledtabular}
\end{table}
%%%%%%%%%%%%%%%%%%%%%%%%%%%%%%%%%%%%%%%%%%%%%%%%%%%%%%%%%%%%%%%%%

%%%%%%%%%%%%%%%%%%%%%%%%%%%%%%%%%%%%%%%%%%%%%%%%%%%%%%%%%%%%%%%%%
\begin{table}
\caption{\label{tab:structure2}
Refined atomic positions for Ba$_3$InIr$_2$O$_9$ at 3.4\,K. The $U_{\rm iso}$ are isotropic atomic displacement parameters (in 10$^{-2}$\,\r A$^2$). Lattice parameters: $a=5.8152(3)$\,\r A, $b=10.0680(5)$\,\r A, $c=14.4619(6)$\,\r A, $\beta=90.854(3)^{\circ}$, $C2/c$, $R_I=0.039$, $R_p=0.061$. Atomic displacement parameters for oxygen were constrained in the refinement.
}
\begin{ruledtabular}
\begin{tabular}{cccccc}
 & & $x/a$ & $y/b$ & $z/c$ & $U_{\rm iso}$ \\
Ba1 & $4e$ & 0 & 0.0010(15) & 0.25 & 1.0(1) \\
Ba2 & $8f$ & 0.0049(9) & 0.3360(11) & 0.0889(3) & 0.1(1) \\
In  & $4a$ & 0 & 0 & 0 & 0.2(1) \\
Ir  & $8f$ & $-0.0081(5)$ & 0.3334(6) & 0.8397(2) & 1.0(1) \\
O1  & $4e$ & 0 & 0.4918(10) & 0.75 & 1.2(1) \\
O2 & $8f$ & 0.230(2) & 0.2573(8) & 0.7541(4) & 1.2(1) \\
O3 & $8f$ & $-0.018(2)$ & 0.1717(8) & 0.9140(6) & 1.2(1) \\
O4 & $8f$ & 0.226(1) & 0.4158(9) & 0.9251(4) & 1.2(1) \\
O5 & $8f$ & $-0.263(1)$ & 0.4144(9) & 0.9079(5) & 1.2(1) \\
\end{tabular}
\end{ruledtabular}
\end{table}
%%%%%%%%%%%%%%%%%%%%%%%%%%%%%%%%%%%%%%%%%%%%%%%%%%%%%%%%%%%%%%%%%

The low-temperature monoclinic distortion is primarily related to the tilting of the IrO$_6$ and InO$_6$ octahedra. It has nearly no effect on relative positions of the Ir atoms. For example, the Ir--Ir distance within the dimer shrinks from 2.637(2)\,\r A at 300\,K to 2.599(4)\,\r A at 3.4\,K, presumably due to thermal expansion. The Ir--Ir distances between the dimers do not change at all, compare 5.832(1)\,\r A at 300\,K to 5.813(8)\,\r A and 5.815(4)\,\r A at 3.4\,K for the nearest-neighbor Ir--Ir distances in the $ab$ plane. Given this negligible structural effect and the absence of any signatures in thermodynamic measurements, we conclude that the hexagonal-to-monoclinic phase transition should be an effect of structural (geometrical) origin and bears no relation to the magnetism of Ba$_3$InIr$_2$O$_9$. Determination of the exact transition temperature requires further dedicated diffraction experiments at intermediate temperatures and lies beyond the scope of our present study.

\subsection{Magnetic susceptibility}
\begin{figure}
\includegraphics[width=0.48\textwidth]{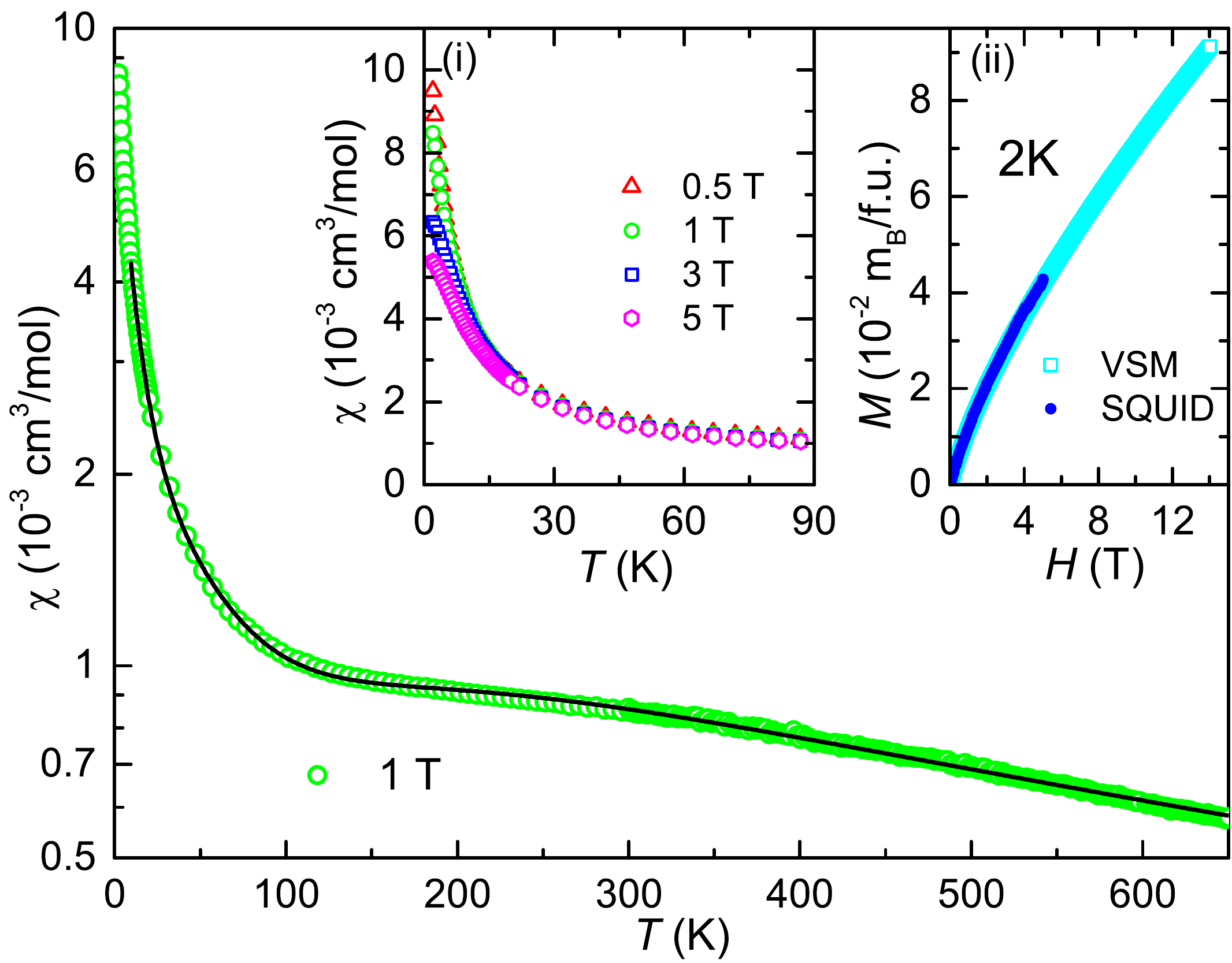}
\caption{\label{fig:Susceptibility}
Magnetic susceptibility of Ba$_{3}$InIr$_{2}$O$_{9}$ at 1\,T is shown in semi-log scale in the temperature range $2-650$\,K. The black solid line is the fit of the data with Eq.~\eqref{eq:susceptibilityfit}. Insets: (i) Magnetic susceptibility of Ba$_3$InIr$_2$O$_9$ shown as function of temperature at various applied fields. (ii) Isothermal magnetization curve measured at 2\,K in SQUID (up to 5\,T) and VSM (up to 14\,T).} 
\end{figure}

Temperature dependence of the magnetic susceptibility ($\chi=M/H$) measured in various applied fields is shown in the inset (i) of Fig.~\ref{fig:Susceptibility}. Down to 2\,K, we did not observe any anomaly or divergence of field-cooled and zero-field-cooled data, suggesting the absence of long-range ordering and spin freezing.

The susceptibility follows the Curie-Weiss (CW) behavior between 10 and 70\,K. At higher temperatures (see Fig.~\ref{fig:Susceptibility}), the data deviate from the CW law due to thermal changes in the mixed-valence Ir$_2$O$_9$ dimers that can adopt different electronic configurations ~\cite{streltsov2016}. A tentative van Vleck fit with the three-level model 
\begin{align}
\chi(T)=&\,\chi_{0}+\frac{N_A\mu_{B}^{2}}{3k_{B}(T-\theta)}\times\frac{3}{4}\times \notag\\
 &\times\frac{g_0^2+5g_{1}^{2}\,e^{-\Delta_{1}/k_{B}T}+5g_{2}^{2}\,e^{-\Delta_{2}/k_{B}T}}{1+e^{-\Delta_{1}/k_{B}T}+e^{-\Delta_{2}/k_{B}T}},
\label{eq:susceptibilityfit}\end{align}
where $N_A$ is Avogadro's number, $g_0$ is electronic $g$-factor for the ground state, and $g_1$ and $g_2$ are $g$-factors for the excited states, which are separated from the ground state by energy gaps of $\Delta_1$ and $\Delta_2$, respectively, yields decent description of the magnetic susceptibility up to at least 650\,K. This fitting function can be understood as follows. In the absence of electronic correlations and spin-orbit coupling, 9 electrons occupying 6 $t_{2g}$ orbitals of two Ir atoms give rise to $S=\frac12$ and $S=\frac32$ states depending on the filling of the molecular orbitals of the dimer~\cite{ziat2017}. The effect of spin-orbit coupling is taken into account by introducing electronic $g$-factors as fitting parameters. Additionally, we assumed that the $S=\frac32$ state splits into two, because fitting with one excited state was not successful, whereas two distinct excited states provide a good description of the susceptibility in the broad temperature range.

Fitting the susceptibility with Eq.~\eqref{eq:susceptibilityfit} yields $\chi_0=-6.8\times 10^{-5}$\,cm$^3$/mol, $\theta=-6.8$\,K, $\Delta_1=107$\,K, $\Delta_2=472$\,K, $g_0=0.872$, $g_1=0.678$, and $g_2=1.810$. The relatively low value of $\Delta_1$ explains the deviation from the Curie-Weiss behavior already above 70\,K. The obtained $\Delta_1$ and $\Delta_2$ are of the same order of magnitude as in the mixed-valence ruthenates isostructural to Ba$_3$InIr$_2$O$_9$~\cite{ziat2017}. The paramagnetic effective moments are $\mu_{\rm eff}=0.76$\,$\mu_B$ in the ground state, $\mu_{\rm eff,1}=1.31$\,$\mu_B$ in the first excited state, and $\mu_{\rm eff,2}=3.51$\,$\mu_B$ in the second excited state. 

%%%%%%%%%%%%%%%%%%%%%%%%%%%%%%%%%%%%%%%%%%%%%%%%%%%%%%%%%%%%%%%%%
%%%%%%%%%%%%%%%%%%%%%%%%%%%%%%%%%%%%%%%%%%%%%%%%%%%%%%%%%%%%%%%%%%%%

For the rest of this paper, we focus on the low-temperature regime below 70\,K with the paramagnetic effective moment of $\mu_{\eff}=0.76$\,$\mu_B$/dimer, which is comparable to the values reported for mixed-valence iridates earlier~\cite{miiller2012,terzic2015}. Antiferromagnetic couplings between magnetic moments localized on the mixed-valent dimers are confirmed by the negative Curie-Weiss temperature $\theta_{\chi}=-7$\,K. Isothermal magnetization curve at 2\,K (see inset (ii) of Fig.~\ref{fig:Susceptibility}) does not show any sign of saturation up to 14\,T. This weak sensitivity to the field may be due to the strongly reduced $g$-factor (only 44\% of its spin-only value) that lessens the effect of the external field.

%%%%%%%%%%%%%%%%%%%%%%%%%%%%%%%%%%%%%%%%%%%%%%%%%%%%%%%%%%%%%%%%%
\begin{figure}
\includegraphics[width=0.50\textwidth]{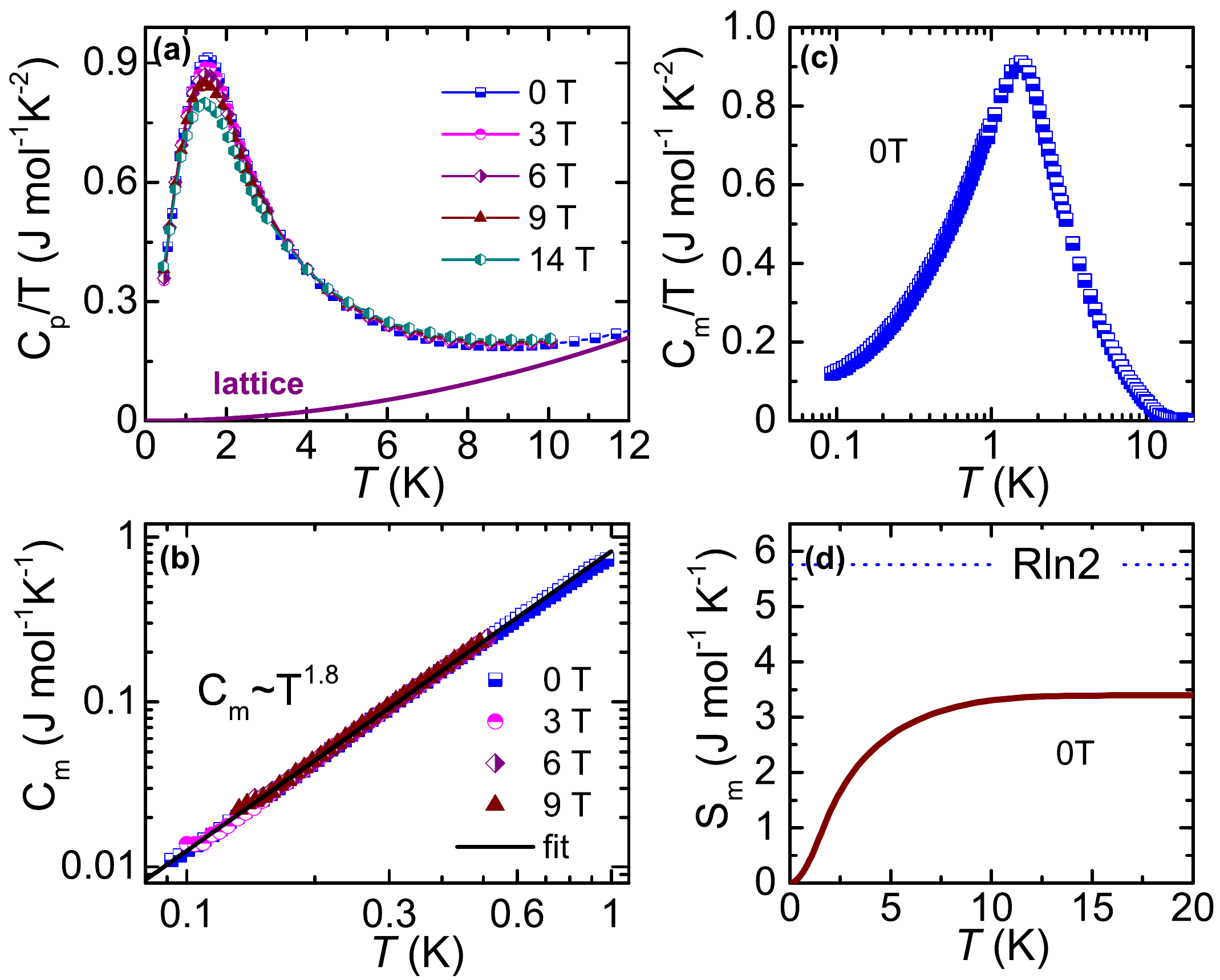}
\caption{\label{fig:Heatcapacity}
(a) Temperature dependence of $C_{p}/T$ at various fields together with the lattice contribution. (b) The low temperature magnetic heat capacity at different fields are shown in log-log scale. The solid line indicates the power law. (c) The temperature dependence of the zero field magnetic heat capacity divided by temperature ($C_{m}/T$) is shown in semi log scale. (d) The entropy change (S$_{m}$) at zero field is shown.} 
\end{figure}
%%%%%%%%%%%%%%%%%%%%%%%%%%%%%%%%%%%%%%%%%%%%%%%%%%%%%%%%%%%%%%%%%%%%

\subsection{Specific heat}
First insight into the low-temperature magnetism is obtained from the specific-heat data. A broad peak in $C_p/T$ is observed around 1.6\,K (Fig.~\ref{fig:Heatcapacity}a) indicative of a crossover between the paramagnetic (thermally disordered) and spin-liquid (quantum disordered) regimes~\cite{li2015,li2016}, further confirmed by the increase in the muon zero-field relaxation rate around the same temperature (Sec.~\ref{sec:musr}). The peak shifts toward lower temperatures in the applied field, although the changes are relatively small even at 14\,T, and no transition anomaly is observed.

Heat capacity was further measured at temperatures well below the broad maximum. For an insulating material, total heat capacity is a sum of the magnetic, lattice, and nuclear contributions, $C_{p}=C_{m}+C_{\rm lat}+C_{\rm nuc}$. To extract the magnetic specific heat of the material, we need to subtract the lattice part and nuclear part from the total specific heat $C_{p}$. In the absence of a suitable non-magnetic analog, we fitted the measured specific heat with $C_{p}=\beta T^{3}$ in the range $14-20$\,K~\cite{supplement}. The fitting of the lattice part yields $\beta=1.45$\,mJ\,mol$^{-1}$\,K$^{-4}$ and the Debye temperature $\Theta_{D}=272$\,K. The fitted curve is extrapolated to low temperatures and taken as the lattice part ($C_{\rm lat}$) that was subtracted from the experimental data. 

At low temperatures, the nuclear contribution becomes prominent. To extract the magnetic heat capacity, we adopted the following procedure:

(1) We fitted the $C_{p}$ data for each field ($B$) from lowest $T$ up to $300-400$\,mK with $C_{p} = \alpha/T^{2} + C_{0}\,T^\gamma$~\cite{supplement}, where $\alpha/T^2$ stands for $C_{\rm nuc}$, and power-law behavior of $C_{\rm mag}$ is assumed.

(2) For each field, the magnetic heat capacity $C_m$ is obtained as $C_m=C_p-C_{\rm nuc}$, whereas $C_{\rm lat}$ is negligible in this temperature range.

(3) Reliability of the $\alpha$ values is verified by plotting the field dependence $\alpha(B^2)$, which was linear, as expected~\cite{supplement}.

%The power-law behavior of $C_m$ is illustrated in Fig.~\ref{fig:cp-gamma}. The value of $\gamma$ remains field-independent up to 9\,T and slightly decreases in higher fields concomitant with the suppression of the 1.6\,K maximum in $C_p/T$ (see Fig.~2b of the manuscript).

After subtracting the nuclear contribution~\cite{supplement}, we arrive at the robust power-law behavior $C_{m}=C_0T^{\gamma}$ with $C_0=832$\,mJ\,mol$^{-1}$\,K$^{-2.83}$ and $\gamma=1.83$ (Fig.~\ref{fig:Heatcapacity}b). This behavior gives first indication of a gapless ground state, because otherwise low-energy excitations over a spin gap would give rise to the exponential decay of $C_m$ at low temperatures. The power-law behavior persists up to at least 14\,T, but above 9\,T the exponent $\gamma$ is slightly reduced~\cite{supplement}.

%%%%%%%%%%%%%%%%%%%%%%%%%%%%%%%%%%%%%%%%%%%%%%%%%%%%%%%%%%%%%%%%%%%%
\begin{figure}
{\centering {\includegraphics[width=0.55\textwidth]{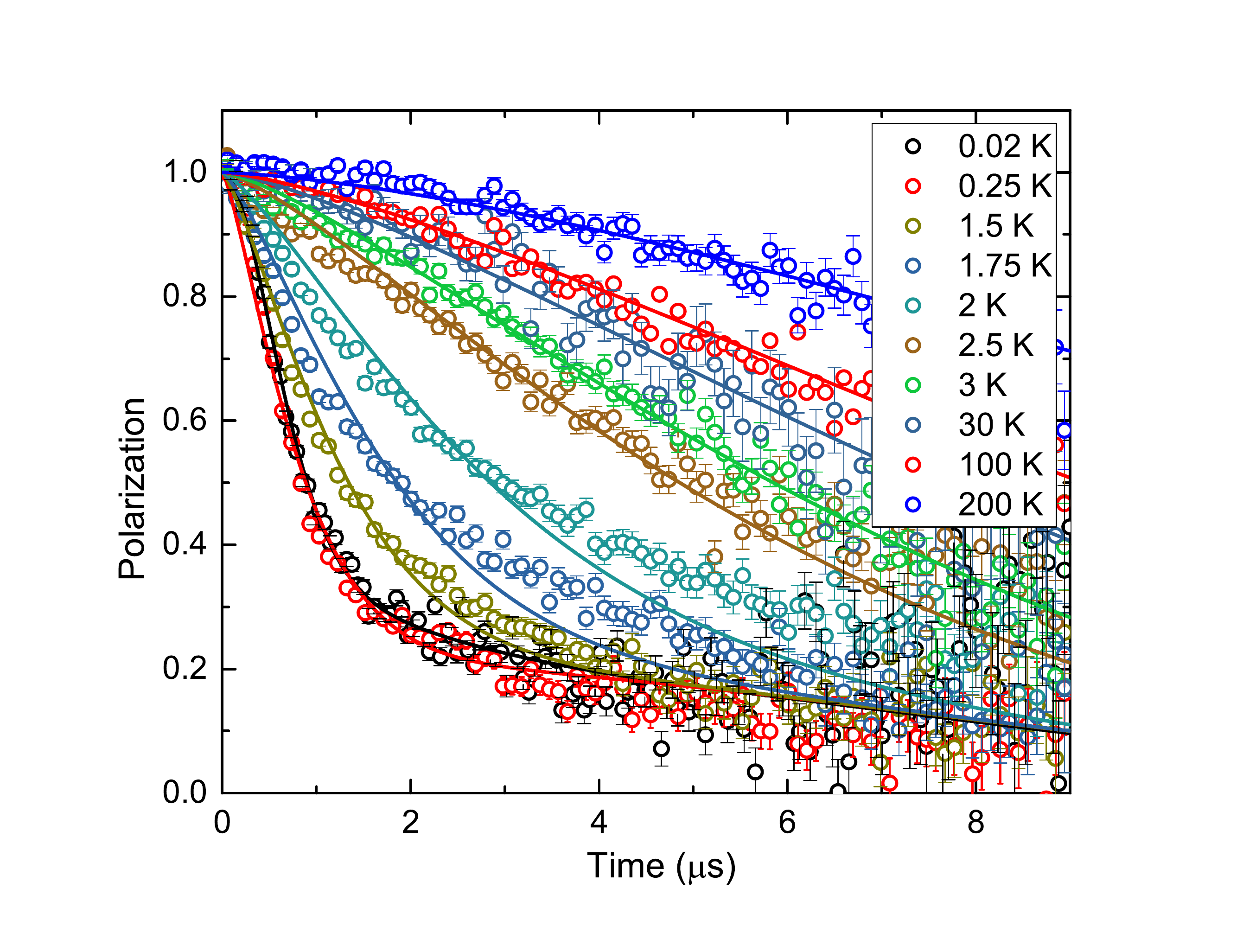}}

} \caption{\label{fig:ZFPolarization}
Muon polarization curves at various temperatures measured under zero field (ZF) along with their fit with Eq. ~\ref{eq:MuonPolarization}.}
\end{figure}
%%%%%%%%%%%%%%%%%%%%%%%%%%%%%%%%%%%%%%%%%%%%%%%%%%%%%%%%%%%%%%%%%%%%

By integrating $C_{m}/T$ in zero field (see Fig.~\ref{fig:Heatcapacity}c) within the temperature range from 0.08 to 20\,K, we estimated magnetic entropy change of 3.4\,J\,mol$^{-1}$\,K$^{-1}$ (see Fig.~\ref{fig:Heatcapacity}d), which is about $60\%$ of the entropy expected for spin-$\frac12$. Part of the magnetic entropy should be then released at higher temperatures, where the magnetic contribution is concealed behind a much larger lattice term. We note here that a two-step release of the magnetic entropy (and, consequently, two well-separated peaks of the magnetic specific heat $C_m$) are not uncommon in frustrated magnets~\cite{sindzingre2000,nasu2015}.

\subsection{$\mu$SR}
\label{sec:musr}
Whereas thermodynamic measurements provide first hints towards the absence of long-range magnetic order in Ba$_3$InIr$_2$O$_9$, experimental evidence for the QSL formation is not complete without a local probe. To this end, we use muon spin relaxation ($\mu$SR), which is a very sensitive technique to detect static local fields arising from weak long-range order or spin freezing. 

The relaxation curves of the muon polarization in zero field (ZF) are shown in Fig.~\ref{fig:ZFPolarization}. The absence of oscillations in the zero-field (ZF) signal and the lack of the polarization recovery to $1/3$ indicate the absence of any frozen moments in Ba$_3$InIr$_2$O$_9$ in the temperature range from $200$\,K down to $20$\,mK. The polarization curves can be fitted with Eq.~\ref{eq:MuonPolarization} which is a combination of the depolarization due to the muon coupled to the In nuclear magnetism (Kubo-Toyabe Gaussian function) and depolarization due to the electronic magnetism evolving with temperature.
\begin{equation}
P(t)=fe^{-(\lambda_{\rm F} t)^{\beta}}+(1-f)\left(\frac{1}{3}+\frac{2}{3}\left(1-(\sigma t)^{2}\right)\,e^{-\frac{(\sigma t)^{2}}{2}}\right)\label{eq:MuonPolarization}
\end{equation} 
Here, $f=0.786(2)$ is the fraction of muons coupled to the electronic magnetism, $\sigma=0.094(2)$\,$\mu$s$^{-1}$ is the nuclear depolarization rate, $\beta=1.308(7)$ is the stretched exponent, and $\lambda$ is the electronic depolarization rate. The stretched exponent $\beta$ is independent of field and temperature. The small deviation from unity could be due to a small distribution of the muon sites close to the In ions.

\begin{figure}
{\centering {\includegraphics[width=0.5\textwidth]{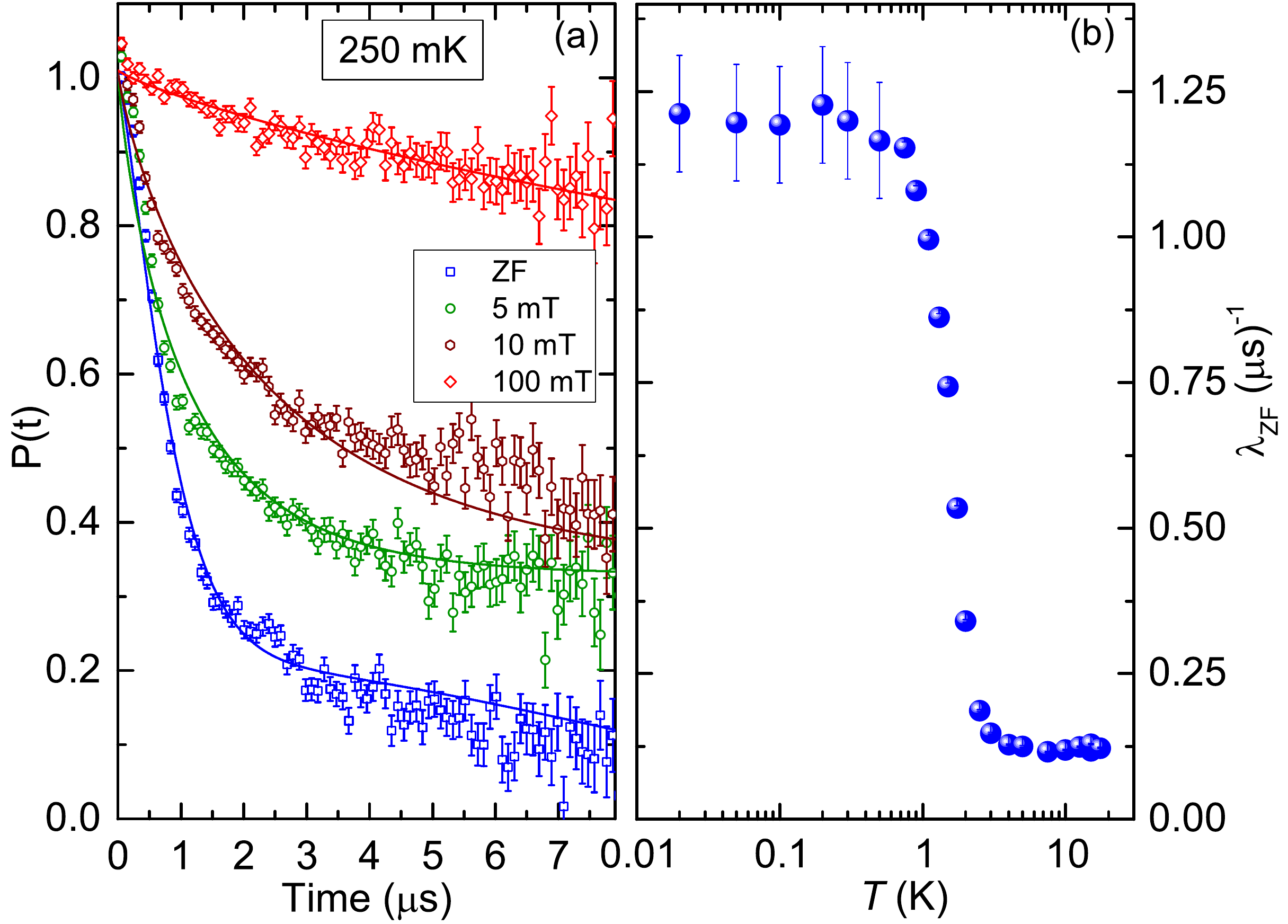}}

} \caption{\label{fig:muSR}
(a) Muon polarization curve at $250$\,mK with various applied magnetic fields are shown along with their fitting (see text). (b) Temperature dependence of $\lambda_{\rm ZF}$ at zero applied magnetic field.}
\end{figure}

The ZF $\mu^{+}$ relaxation rate ($\lambda_{\rm ZF}$) obtained from fitting the ZF muon depolarization curves is shown as function of temperature in Fig.~\ref{fig:muSR}b. At high temperatures ($20$\,K to $3$\,K), $\lambda_{\rm ZF}$ remains constant at $\sim0.12\thinspace\mu$s$^{-1}$, which is consistent with the paramagnetic fluctuations of Ir moments
according to the Bloembergen Purcell, and Pound theory~\cite{pound1948}.

From $3$\,K down to $1$\,K, $\lambda_{\rm ZF}$ increases with decreasing temperature. The enhancement of $\lambda_{\rm ZF}$ in a narrow temperature window indicates a slowing down of Ir spin fluctuations due to the development of strong short-range correlations, a common feature seen in other QSL candidates \cite{fak2012,clark2013,li2016}. Upon further cooling, $\lambda_{\rm ZF}$ shows temperature-independent plateau-like behavior between $1$\,K and $20$\,mK. The plateau-like behavior in $\lambda_{\rm ZF}$ vs $T$ (see Fig.~\ref{fig:muSR}b) has also been observed in several other QSL compounds. 

To verify if the plateau-like behavior is originating from the muons directly coupled to the frustrated spins \cite{colman2011} or the muons coupled to defects \cite{kermarrec2011,gomilsek2016}, we have performed $\mu$SR experiments under transverse field (TF) of 0.4\,T and estimated the $\mu$SR line shift ($K^{\mu}$) as a function of temperature ~\cite{orain2014} as shown in Fig.~\ref{fig:NMR}a. $K^{\mu}$ increases with decreasing temperature and saturates below $3$\,K, from where $\lambda$ (see Fig.~\ref{fig:muSR}b) starts increasing. This indicates that antiferromagnetic (AFM) spin fluctuations are dominating at low temperature in Ba$_3$InIr$_2$O$_9$. The similarity between the temperature dependence of $K^{\mu}$ and NMR shift ($K_{\rm NMR}$) (see Fig.~\ref{fig:NMR}a) proves that the muons are directly coupled to the Ir moments. 

%%%%%%%%%%%%%%%%%%%%%%%%%%%%%%%%%%%%%%%%%%%%%%%%%%%%%%%%%%%%%%%%%%%%
%%%%%%%%%%%%%%%%%%%%%%%%%%%%%%%%%%%%%%%%%%%%%%%%%%%%%%%%%%%%%%%%%%%%

In order to identify the origin of the depolarization rate being static or dynamic, we performed decoupling experiments at $250$\,mK (shown in Fig.~\ref{fig:muSR}a) applying longitudinal fields (LF) from $5$\,mT up to $100$\,mT. Should the low-temperature plateau in $\lambda_{\rm ZF}$ arise from a static field, the size of this field can be estimated as $B_{\rm loc}=\lambda/\gamma_{\mu}\approx1.4$\,mT, where $\gamma_{\mu}=135.5\times2\pi$\,s$^{-1}$$\mu$T$^{-1}$ is the gyromagnetic ratio for muons. The static field can then be decoupled by applying an external magnetic field $\simeq14$\,mT, which is one order of magnitude higher than $B_{\rm loc}$. The polarization curve under a field of $100$\,mT does not show the signs of a full polarization,
suggesting that the plateau does not originate from a static internal field and the spins are dynamic in nature even at the lowest temperature of $20$\,mK, as expected in a QSL. The plateau-like behavior of $\lambda_{\rm ZF}$ toward lowest temperature indicates much slower spin dynamics of the material compared to the $\mu$SR time window. It is another characteristics of QSL candidates~\cite{fak2012,clark2013,li2016}.

%%%%%%%%%%%%%%%%%%%%%%%%%%%%%%%%%%%%%%%%%%%%%%%%%%%%%%%%%%%%%%%%%%%%
\begin{figure}
{\centering {\includegraphics[width=0.46\textwidth]{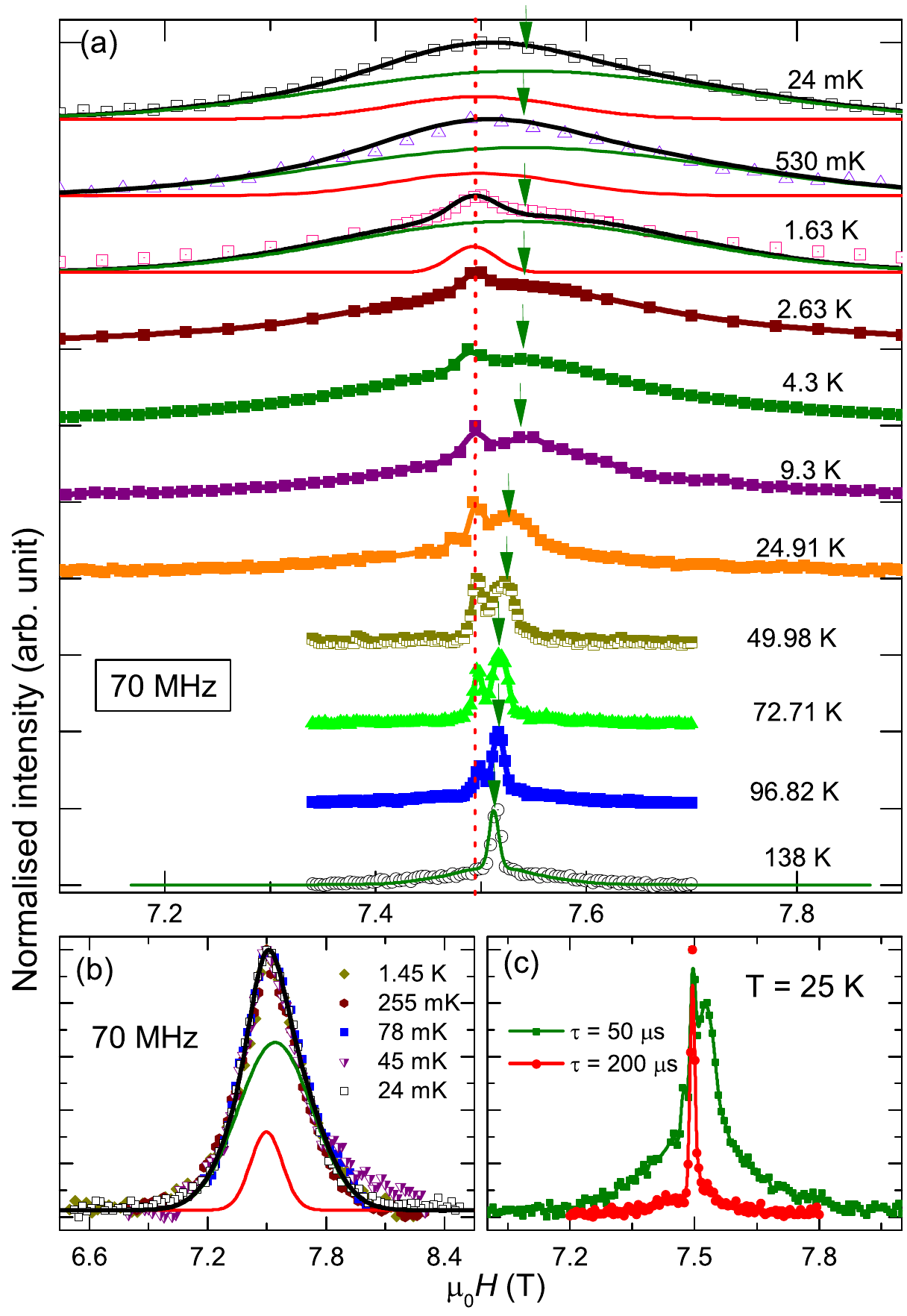}}

} \caption{\label{fig:Spectra}
(a) Temperature evolution of NMR spectra measured
at $70$\,MHz is shown. Individual spectra are shifted along y-axis
for clarity. The vertical dotted red line indicates the diamagnetic
resonance field as a reference. The green and red solid lines represent the intrinsic and impurity contributions, respectively, in the total spectra (black solid line). The green arrow marks the peak position of the intrinsic line. (b) Normalized $^{115}$In NMR spectra at different temperatures between $1.45$\,K down to $24$\,mK. The green, red, and black solid lines correspond to the spectra at $24$\,mK. (c) NMR spectra measured with different pulse separation ($\tau$) at $25$\,K.}
\end{figure}
%%%%%%%%%%%%%%%%%%%%%%%%%%%%%%%%%%%%%%%%%%%%%%%%%%%%%%%%%%%%%%%%%%%%

\subsection{NMR}
Nuclear magnetic resonance (NMR) is a concurrent probe for local magnetic fields and spin dynamics. $^{115}$In NMR spectra measured at $70$\,MHz in the temperature range $0.024-138$\,K are shown in Fig.~\ref{fig:Spectra}(a). At high temperatures (for example, at $138$\,K), the spectra exhibit one single isotropic line which indicates the unique crystallographic site for indium, consistent with our structural analysis. The spectra can be described satisfactorily assuming the $I=9/2$ nuclei with the $380$\,kHz quadrupolar coupling constant ($\nu_{Q}$). 

Below about $100$\,K, we observed a second line, which remained unshifted within the entire temperature range. This second line presumably originates from trace amounts (0.9\,\%) of a non-magnetic impurity phase present in the sample. For estimating the impurity contribution in the spectral line shape, we measured the spectra at several temperatures with $\tau=50$\,$\mu$s and $\tau=200$\,$\mu$s, where $\tau$ is the pulse separation between the spin-echo sequence $\pi/2-\tau-\pi$. The intrinsic line has much shorter spin-lattice relaxation time $T_{1}$ as well as the spin-spin relaxation time $T_{2}$ compared to the impurity line~\cite{supplement}. Hence it is expected that with longer $\tau$ the intrinsic part of the spectra is already relaxed, whereas the impurity contribution sustains. The comparison of the spectra corresponding to different $\tau$ measured at $T=25$\,K is shown in Fig.~\ref{fig:Spectra}c.

\begin{figure}
\includegraphics[width=0.48\textwidth]{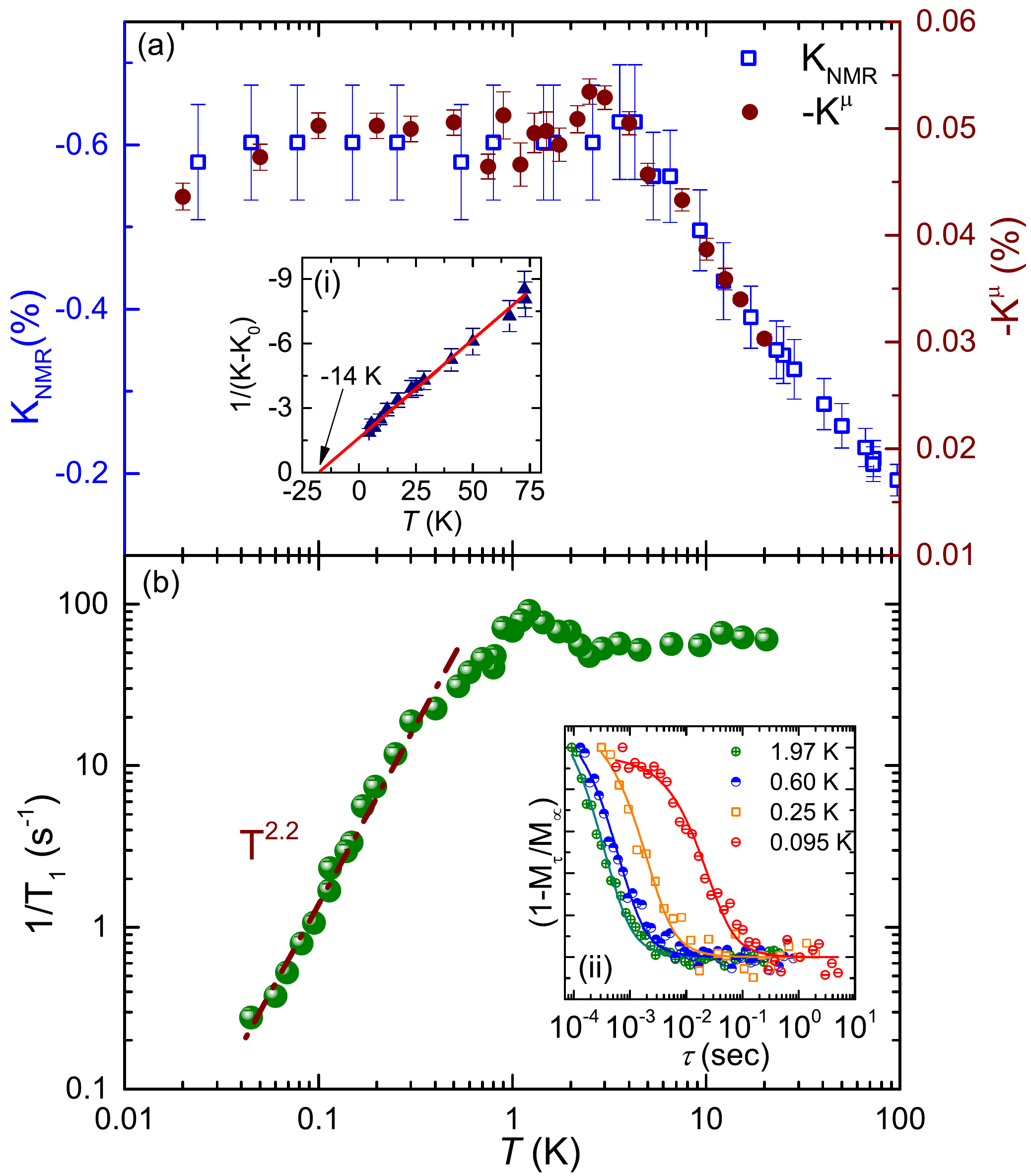}
\caption{\label{fig:NMR}
(a) Temperature dependence of the NMR ($^{115}K$)  and $\mu$SR ($-K^{\mu}$) line shift are shown corresponding to the left and right axes, respectively. (b) Temperature dependence of $1/T_{1}$. The dashed line is a guide-to-the-eye for the power-law $\propto T^{2.2}$ behavior. Insets: (i) $1/(K-K_{0})$ as function of temperature with the linear CW fit. (ii) The nuclear magnetization recoveries curves along with their fitting at different temperatures.}
\label{structure}
\end{figure}

The spectra are described at all temperatures below 100\,K with two Gaussian lines, one for magnetic (intrinsic) and the other one for a non-magnetic (impurity) contribution as shown in Fig.~\ref{fig:Spectra}(a)). At low temperatures, the two lines (intrinsic line and the impurity line) merge together. For estimating the Knight shift of the intrinsic line, we have used the peak of the intrinsic contribution marked with the green arrows in Fig.~\ref{fig:Spectra}a. This peak shifts towards higher fields with decreasing temperature yielding the local spin susceptibility $K$. In general,

\begin{equation}
 K=K_{0}+(A_{\rm hf}/N_{A}\mu_{B})\,\chi(T),
\end{equation}
where $K_{0}$ is the temperature-independent part of the line shift $K$, and $N_{A}$ is the Avogadro's number. The hyperfine coupling constant is estimated at $A_{\rm hf}=-1.675$\,T/$\mu_{B}$~\cite{supplement}. The temperature dependence of the $^{115}$In line shift ($K$) is shown in Fig.~\ref{fig:NMR}a. A continuous increase in the line shift $K(T)$ with decreasing temperature from $100$\,K down to $\sim4$\,K indicates the development of spin correlations. The inverse of the temperature-dependent part of the line shift $1/(K-K_{0})$ is shown as a function of temperature in the inset of Fig.~\ref{fig:NMR}a. A CW fit of the data yields the CW temperature $\theta_{\rm NMR}=-14$\,K. This value should be more reliable than $\theta_{\chi}=-7$\,K estimated from the bulk susceptibility, since the NMR line shift does not contain any impurity contributions.

At low temperature the total spectra remain unchanged as shown in Fig ~\ref{fig:Spectra}b. This indicates that both the line shift $K$ and line width $\Delta H$ of the spectra remain constant down to the base temperature of $25$\,mK. Saturation of both these quantities suggests saturation of spin correlations below $\sim4$\,K. The finite and temperature-independent value of $K$ below 4\,K gives strong evidence for the gapless behavior. From $\Delta H$, we estimated the saturated magnetic moment $\mu_{s}=0.89$\,$\mu_{B}$/f.u. close to 1\,$\mu_B$ expected for one unpaired electron per the Ir$_2$O$_9$ dimer.

%%%%%%%%%%%%%%%%%%%%%%%%%%%%%%%%%%%%%%%%%%%%%%%%%%%%%%%%%%%%%%%%%
%%%%%%%%%%%%%%%%%%%%%%%%%%%%%%%%%%%%%%%%%%%%%%%%%%%%%%%%%%%%%%%%%%%%

To probe the low-energy spin dynamics, we have studied the NMR spin-lattice relaxation rate ($1/T_{1}$) at various temperatures. ($1/T_{1}$) was measured at the peak position of the intrinsic line using long pulses to avoid any interference from the impurity line. $1/T_{1}$ at different temperatures are obtained by fitting (shown in the inset of Fig.~\ref{fig:NMR}b) the longitudinal nuclear magnetization recovery curves with the equation
\begin{align*}
 1-M_{t}/M_{\infty} =& C\,( 0.006\,e^{-(2Wt)^{\beta}}+0.0335\,e^{-(12Wt)^{\beta}}+ \\ &+0.0925\,e^{-(30Wt)^{\beta}}+0.215\,e^{-(56Wt)^{\beta}} \\ &+0.653\,e^{-(90Wt)^{\beta}}),
\end{align*}
where $1/T_{1}=2W$. Here, $M$ is the nuclear magnetization, $\beta$ is a stretched exponent, which accounts for a distribution of the $T_{1}$ values due to disorder, and $C$ is a pre-factor. 

In general,
\begin{equation}
 (1/T_{1}T)\sim\sum_{q}A_{\rm hf}(q)^{2}\times\chi^{\prime\prime}(q,\omega\rightarrow0),
\end{equation} where $\chi^{\prime\prime}(q,\omega)$ is the imaginary part of the dynamical spin susceptibility. The temperature dependence of $1/T_{1}$ is shown in Fig.~\ref{fig:NMR}b. Between $30$\,K and $4$\,K, $1/T_{1}$ is almost temperature-independent, indicating the paramagnetic behavior. A weak hump-like feature is seen around $1.6$\,K, which coincides with the hump in the $C_{p}/T$ data (Fig.~\ref{fig:Heatcapacity}a) and with the rapid increase in $\lambda_{\rm ZF}$ (Fig.~\ref{fig:muSR}b). No long-range order occurs in Ba$_3$InIr$_2$O$_9$. Therefore, this hump is not a broadened signature of a magnetic transition. Instead, it may indicate some intrinsic physics, as recently proposed for the honeycomb iridates~\cite{yoshitake2016}. The applicability of this scenario to Ba$_3$InIr$_2$O$_9$ may be an interesting venue for future research.

Below 0.3\,K, ($1/T_1$) follows the $\sim T^{2.2}$ power law similar to the $\sim T^{1.83}$ behavior of $C_m(T)$.

\section{Discussion and Summary}
Our results establish gapless ground state and persistent spin dynamics down to at least 20\,mK, rendering Ba$_3$InIr$_2$O$_9$ a potential QSL material. However, its low-temperature behavior is rather different from theoretical expectations for a gapless QSL, where magnetic excitations are usually described in terms of a spinon Fermi surface and should give rise to the linear behavior of both NMR spin-lattice relaxation rate and magnetic specific heat~\cite{lawler2008,yamashita2008,yamashita2011} at low temperatures. Our data for Ba$_3$InIr$_2$O$_9$ clearly deviate from this scenario.

Microscopically, we expect two superexchange pathways. The mixed-valence Ir$_2$O$_9$ dimers may interact in the $ab$ plane via the coupling $J_T$ (the Ir--Ir distance of 5.83\,\r A) and along the $c$ direction via the coupling $J_H$ (the Ir--Ir distance of 5.71\,\r A), see Fig.~\ref{fig:Structure}. Note that we consider Ir--Ir distances, because magnetic electrons are expected to occupy a molecular orbital of the dimer with equal contributions of both Ir sites~\cite{streltsov2016}. While the coupling $J_T$ forms a triangular lattice in the $ab$ plane, as already anticipated in Ba$_3$IrTi$_2$O$_9$~\cite{dey2012,kumar2016,lee2017,becker2015,catuneanu2015}, the coupling $J_H$ leads to a buckled honeycomb geometry. Depending on the ratio between $J_T$ and $J_H$, the system can interpolate between the purely two-dimensional regime ($|J_T|\gg |J_H|$, or vice versa) and a 3D behavior when both $|J_H|$ and $|J_T|$ are of the same size.

On the phenomenological level, quadratic behavior of the zero-field specific heat is expected in an algebraic spin liquid on the kagome lattice~\cite{ran2007,hermele2008}. Another manifestation of the $T^2$ behavior was proposed for a spin model combining the triangular and honeycomb geometries~\cite{chen2012}, as in Ba$_3$InIr$_2$O$_9$, but its applicability to our system is obscured by the large single-ion anisotropy term that is central to the model but seems unlikely in iridates~\cite{winter2016}. Intriguingly, the robust quadratic behavior of the zero-field specific heat has been reported in frustrated magnets with slow dynamics~\cite{ramirez1992,nakatsuji2005} and interpreted in the framework of Halperin-Saslow modes for two-dimensional spin glasses~\cite{sachdev1992,podolsky2009}. It was also observed in Li$_2$RhO$_3$, the Kitaev antiferromagnet, where spins freeze below 6\,K~\cite{khuntia2015}. To what extent these scenarios apply to Ba$_3$InIr$_2$O$_9$ with its clear signatures of persistent spin dynamics should be a subject of future investigations.

Mixed-valence dimers are also reported in ruthenates with the hexagonal perovskite structure. However, these compounds show long-range magnetic order or disordered static magnetism at low temperatures~\cite{ziat2017}. This is different from our case, where persistent spin dynamics is observed. From magnetism point of view, mixed-valence hexagonal perovskites remain a largely uncharted territory. The nature of local moments within the mixed-valence dimers requires further investigation, and peculiarities of magnetic interactions between the dimers remain to be explored. Experiments reveal instances of non-trivial magnetic states, including the potential QSL state in Ba$_3$InIr$_2$O$_9$, and call for a better understanding from theoretical and microscopic perspectives.

Altogether, we demonstrated that Ba$_3$InIr$_2$O$_9$ is a structurally well-ordered QSL candidate showing the \mbox{$\sim T^{1.83}$} behavior of zero-field specific heat and the $\sim T^{2.2}$ power-law for the spin-lattice relaxation rate at low temperatures. Not only would it be interesting for further experimental (probe of the spin excitations) and theoretical (microscopic analysis of the magnetic model) research, it also puts forward mixed-valence iridates as a promising playground for finding new QSL materials. The unique combination of the triangular and buckled honeycomb geometries bears strong connections to current theoretical models of frustrated magnetism~\cite{kitaev2006,rousochatzakis2015,jackeli2015,becker2015,li2015b,rousochatzakis2016}, and may lead to novel manifestations of the QSL physics when different lattice geometries concur.

\acknowledgments
We gratefully acknowledge financial support from the Deutsche Forschungsgemeinschaft (DFG) within the collaborative research center TRR\,80 (Augsburg/Munich). TD, MM, and AT were funded by the Federal Ministry for Education and Research through the Sofja Kovalevskaya Award of Alexander von Humboldt Foundation. FB acknowledges financial support from the project SOCRATE (ANR-15-CE30-0009-01) of the ANR French agency. PK acknowledges support from the European Commission through Marie Curie International Incoming Fellowship (PIIF-GA-2013-627322). Part of the work was performed at the LTF and DOLLY spectrometers of the Swiss Muon Source (Paul Scherrer Institute, Villigen, Switzerland). We thank C. Baines for his support on LTF, and Dana Vieweg for her help with high-temperature susceptibility measurements.

%\bibliography{Ba3InIr2O9}

\newpage
\begin{widetext}
\begin{center}
\large\textbf{\textit{Supplemental Material}\smallskip \\ Persistent low-temperature spin dynamics in mixed-valence iridate Ba$_{3}$InIr$_{2}$O$_{9}$}
\end{center}
\end{widetext}

\renewcommand{\thefigure}{S\arabic{figure}}
\renewcommand{\thetable}{S\arabic{table}}
\setcounter{figure}{0}
\begin{widetext}

\begin{figure}[h]
\begin{centering}
\includegraphics[scale=0.35]{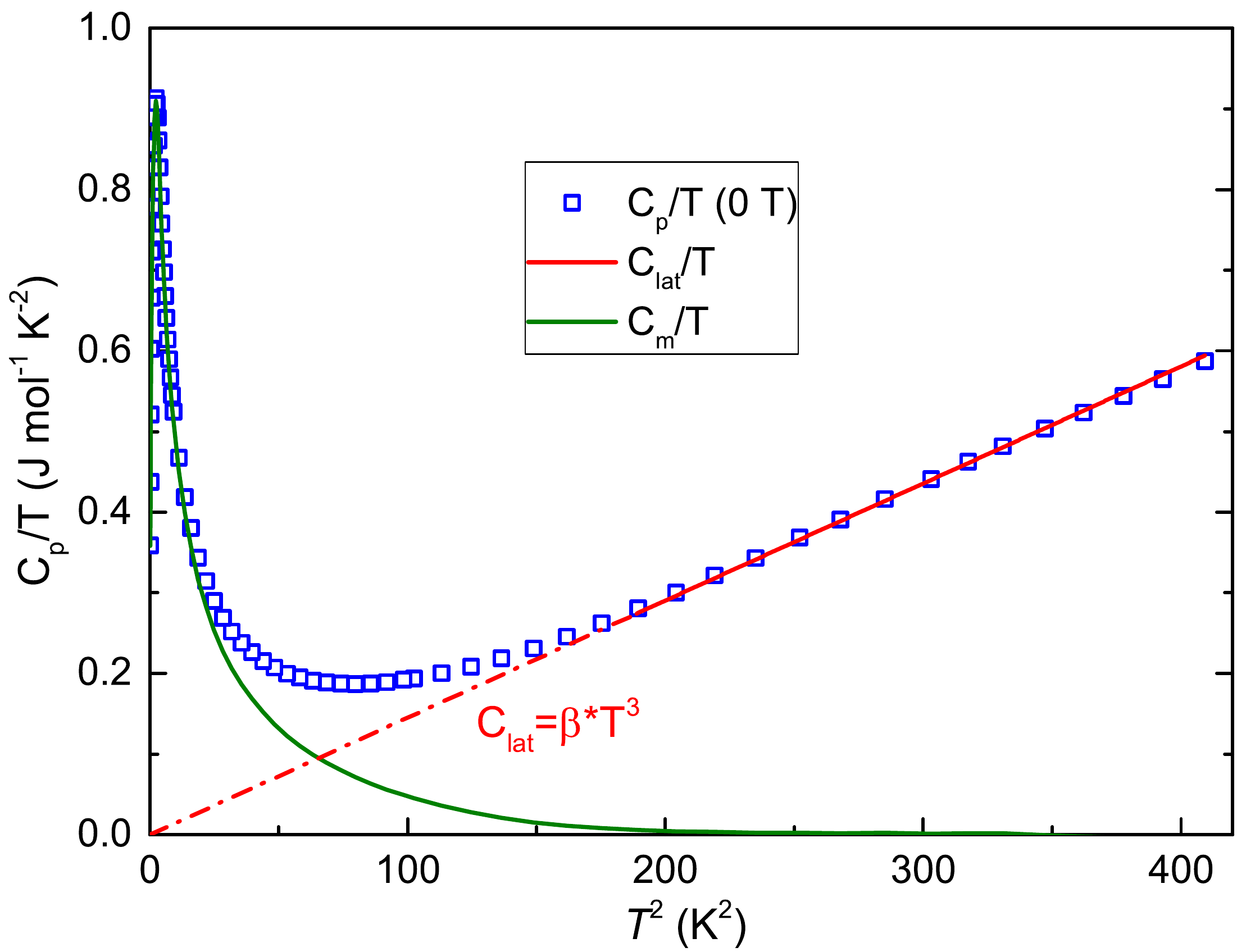}
\par\end{centering}

\centering{}\caption{\label{fig:Clattice} $C_{p}/T$ is shown as function of $T^{2}$
for the zero field data. The linear fit of the data in the range $14-20$\,K
(red solid line) and its extrapolation down to low temperature (red
dashed line) is shown. $C_{m}/T$ obtained after subtracting the lattice
part is shown as green solid line. 
}
\end{figure}

\begin{figure}
\centerline{\includegraphics[scale=0.25]{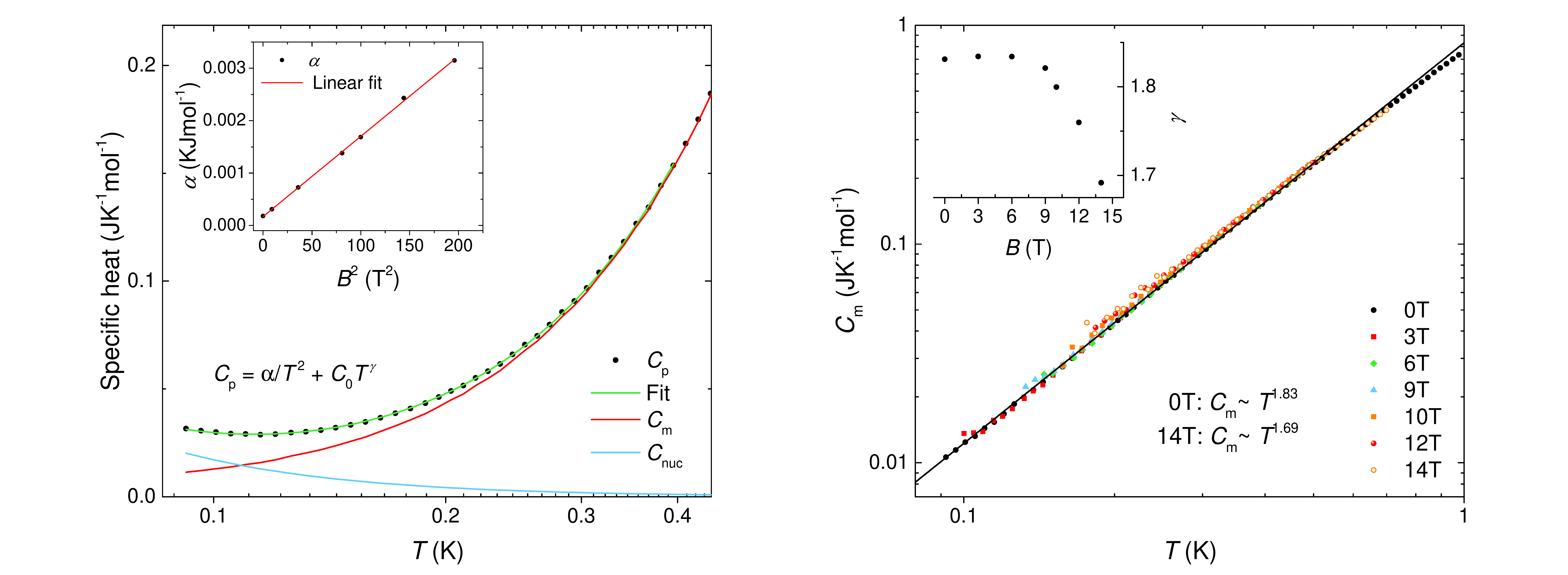}}
\caption{\label{fig:cp-fit}
Left: Determination of the nuclear contribution for the specific heat measured in zero field, see text for details. The inset shows linear dependence of $\alpha$ on $B^2$. Right: Power-law behavior of the magnetic specific heat, $C_m=C_0T^{\gamma}$, in different applied fields. The inset shows field dependence of $\gamma$.
}
\end{figure}

\begin{figure}
\begin{minipage}{8.5cm}
\includegraphics[scale=0.9]{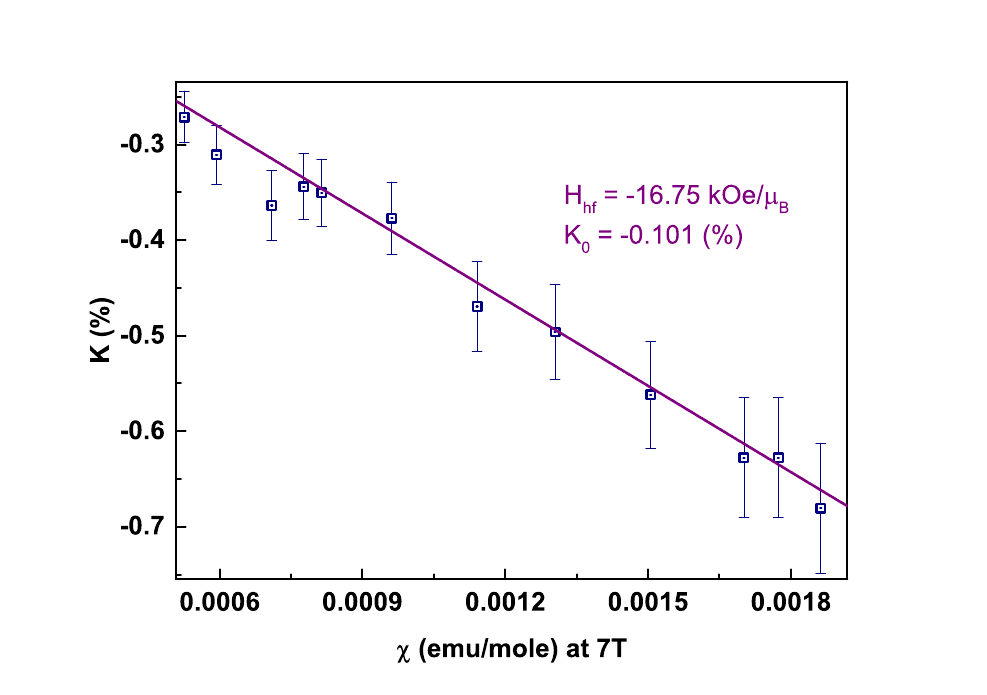}
\end{minipage}
\begin{minipage}{8.5cm}
\includegraphics[scale=0.3]{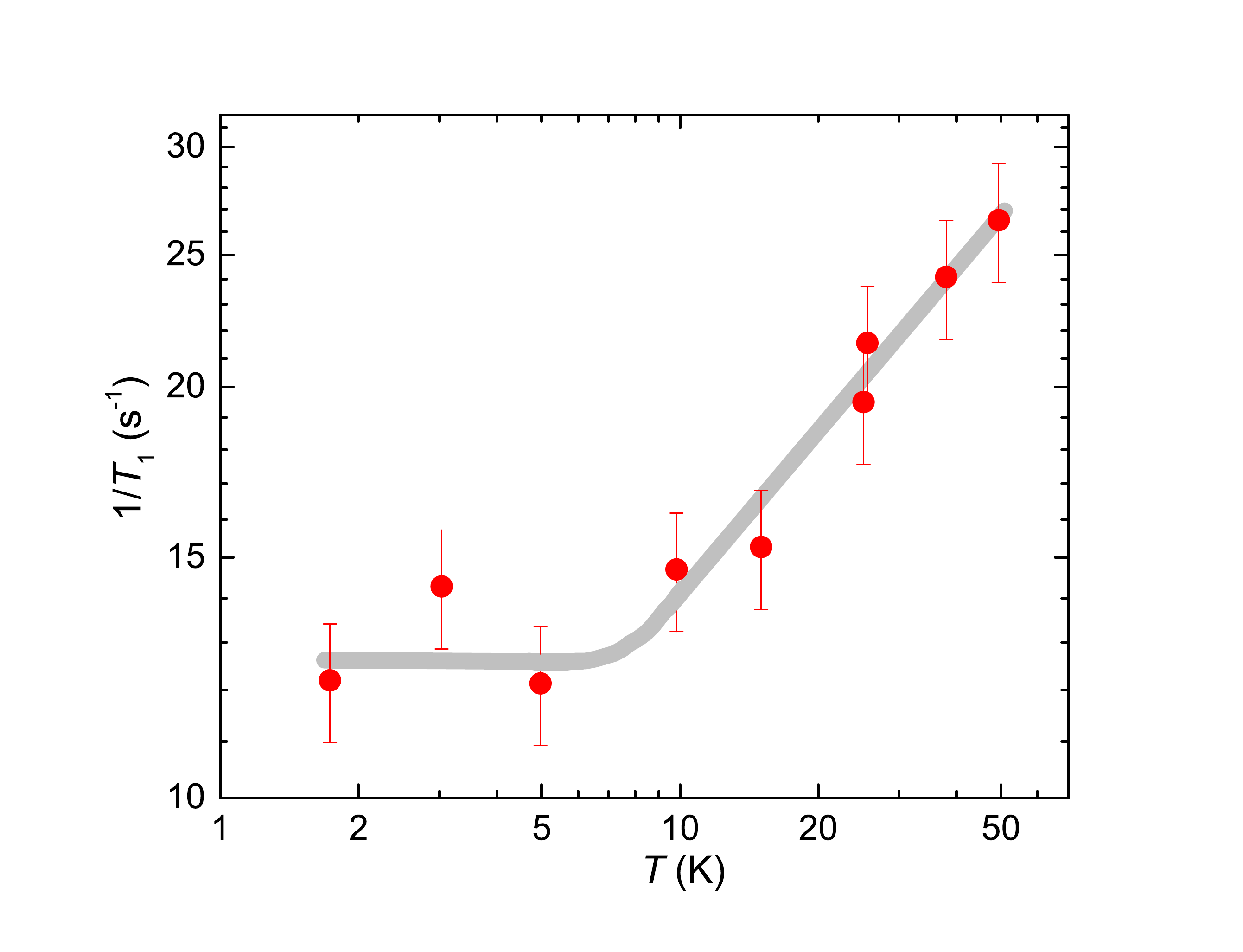}
\end{minipage}
\caption{\label{fig:KvsChi} Left: Knight shift $K$ versus susceptibility $\chi$ with temperature as an implicit parameter. The solid line is a linear fit. Right: Temperature dependence of $1/T_{1}$ for the
non-magnetic impurity phase; the grey line is a guide to the eye. By solely exciting the impurity line, we have measured the relaxation rate of the impurity line as a function of temperature. This relaxation rate decreases towards 10\,K and remains constant at lower temperatures. The relaxation rate is much smaller compared to that of the intrinsic line at elevated temperatures ($T>2$\,K). From the $T_2$ of the impurity line, we estimate the amount of the impurity at 0.9\,\%.}
\end{figure}
\end{widetext}

\end{document}